\documentclass[english,aps,manuscript]{revtex4}
\usepackage[T1]{fontenc}
\usepackage[latin1]{inputenc}
\usepackage{graphicx}
\usepackage{amssymb}

\makeatletter

\providecommand{\tabularnewline}{\\}

\makeatletter

\newcommand{\bee}{\begin{equation}}
\newcommand{\ee}{\end{equation}}
\newcommand{\beea}{\begin{eqnarray}}
\newcommand{\eea}{\end{eqnarray}}

\def\Tr{{\rm tr}}

\makeatother

\usepackage{babel}
\makeatother
\begin{document}

\title{Hypercubic Smeared Links for Dynamical Fermions}

\author{Anna Hasenfratz}

\email{anna@eotvos.colorado.edu}

\affiliation{Department of Physics, University of Colorado, Boulder, CO-80309-390}

\author{Roland Hoffmann}

\email{hoffmann@pizero.colorado.edu}

\affiliation{Department of Physics, University of Colorado, Boulder, CO-80309-390}

\author{Stefan Schaefer}

\email{stefan.schaefer@desy.de}

\affiliation{NIC, DESY, Platanenallee 6, D-15738 Zeuthen, Germany}

\begin{abstract}
We investigate a variant of hypercubic gauge link smearing where the
$SU(3)$ projection is replaced with a normalization to the corresponding
unitary group. This smearing is  differentiable and thus suitable
for use in dynamical fermion simulations using molecular dynamics
type algorithms. We show that this smearing is as efficient as projected
hypercubic smearing in removing ultraviolet noise from the gauge fields.
We test the normalized hypercubic smearing in dynamical improved (clover) 
Wilson and valence overlap simulations. 
\end{abstract}
\maketitle
\tableofcontents{}

\section{Introduction}

In recent years, significant progress has been made in full QCD lattice
simulations. There are simulations with 2+1 flavors, with realistic
quark masses, and in large volumes, though frequently only two of
the three conditions are met at once. These simulations are performed
with different kinds of fermion formulations, from the simplest unimproved
Wilson fermions to highly improved nearly chiral fermions, with improved
rooted staggered fermions and even with the expensive but exactly
chiral overlap fermions. All these calculations, even those with inexact
chiral symmetry, are still expensive and require large computer resources.
Improving the fermionic action such that simulations could be performed
on coarser lattices, or improving the performance of algorithms to
better fit today's computer power is important for truly realistic
simulations. It seems that a simple modification, the use of smeared
gauge fields in the fermionic action, can help improve both the action
and the computational performance as well.

Smeared links are a natural part of improved fermionic actions. In
the perfect action formulation the Dirac operator at the renormalization
group fixed point is fitted by an extended but ultra-local Dirac operator.
This fit is not feasible unless the gauge links of the Dirac operator
are smeared~\cite{Hasenfratz:2001hr}. The exactly chiral overlap
operator~\cite{Neuberger:1997fp} effectively also contains smeared
links, even if the kernel operator is based on thin links. This can
be seen when one considers the expanded form of the overlap formulation
with the square root term in $d(-R_0)/\sqrt{d^{\dagger}(-R_0)d(-R_0)}$.
The order $d^{3}$, $d^{5}$, etc. terms all contribute to the nearest
neighbor fermion coupling of the overlap Dirac operator with extended
gauge connections. The most frequently used staggered fermion formulation,
the so called Asqtad action, also uses fat links~\cite{Orginos:1999cr}.
The smeared links discussed above are part of the definition of the Dirac
operator. The gauge action is independent and in most cases is not smeared.

The effect of smearing is two-fold. First, it averages out small scale
vacuum fluctuations, reducing the non-physical ultra-violet noise
in the fermionic action, secondly it removes extreme local fluctuations
of the gauge fields, lattice dislocations. In the various fermion
discretizations, the effect of vacuum fluctuations and dislocations
comes in different disguises. Staggered fermions' taste breaking is
triggered by gauge field fluctuations within the hypercube and smearing
can effectively reduce this effect~\cite{Orginos:1999cr,Hasenfratz:2001hp}.
For Wilson fermions dislocations  contribute to
the spread of the near zero real modes of the Dirac operator. Those
modes make it impossible to simulate at small quark masses without
going to very fine lattice spacing, and/or large volumes \cite{DelDebbio:2005qa}.
Smearing removes the dislocations and reduces the spread of the eigenmodes~\cite{DeGrand:1998mn}.

Chiral fermions can also benefit from smeared links. The cost of the
overlap operator is largely given by the density of low modes of the
kernel operator from which it is constructed. Smearing reduces the
occurrence of these low modes and thereby can reduce the cost of applying
the operator by an order of magnitude~\cite{DeGrand:2004nq,DeGrand:2005vb}.
In simulations using domain wall fermions, the low modes of the kernel
operator are known to cause explicit breaking of chiral symmetry,
indicated by a non-vanishing residual mass. If there are fewer of
those modes, chiral symmetry is realized to a higher degree and one
can use a smaller fifth dimension without increasing the residual mass.

There is no unique criterion what constitutes a {}``good'' smearing
procedure besides the explicit construction of the fixed point Dirac
operator or the expanded form of overlap fermions. Without such guiding
principles, any smearing, as long as it consists of adding irrelevant
(local) operators to the action, is acceptable. The smeared links
do not even have to be $SU(3)$ elements as is illustrated by the
success of the Asqtad action. Any acceptable procedure will lead to
a valid action, but chosen properly, smearing will improve the scaling
of the continuum limit. If the modification of the gauge fields are
too weak, the smearing has no effect. On the other hand, a definition
of the fat link which spreads over many sites and heavily mixes the
links can lead to an action which again has strong cut-off effects~\cite{DeGrand:2002vu}.
Thus, an optimal smearing is as local as possible while removing as
much of the short scale fluctuations as possible.

The first smearing was introduced by the APE collaboration~\cite{Albanese:1987ds}
and different forms of smearing have been used in quenched studies
since then. Dynamical simulations with smeared links became practical
when the fully differentiable stout smearing was proposed by Morningstar
and Peardon~\cite{Morningstar:2003gk}. Iterating either APE or stout
links can wash out short to intermediate scale \emph{physical} properties
of the action, leading to large scale violations in quantities sensitive
to those scales. Hyper-cubic (HYP) blocking, introduced in Ref.~\cite{Hasenfratz:2001hp},
circumvents this problem by reducing the spread of consecutive smearing
steps. In this paper, we will discuss variants of the HYP blocking
that are differentiable and suitable for molecular dynamics simulations.

In the next Section we first modify the APE construction by replacing
the original $SU(3)$ projection by a normalization to $U(3)$. 
These normalized n-APE links are differentiable and as effective
in removing short scale vacuum fluctuations as the projected APE smearing.
Next we combine n-APE smearing with the HYP definition and show that
n-HYP links are as effective as 3 levels of stout smearing and are
considerably better than HYP links constructed from stout smearing.
The differentiable n-HYP smearing can be used in dynamical simulations
and in Sect. \ref{sec:Force-of-the} we give details of how the fermionic
force can be evaluated with n-HYP smearing. This force term can be
combined with any fermionic action and in Sect. \ref{sec:Numerical-tests}
we illustrate the effectiveness of the smearing both with overlap
and Wilson clover fermions.

\section{Definition of the smeared links}

The APE smeared link~\cite{Albanese:1987ds} is  the basis of
most smearing methods.
First the staple sum $\Gamma_{n,\mu}=\sum_{\nu\neq\mu}U_{n,\nu}U_{n+\nu,\mu}U_{n+\mu,\nu}^{\dagger}$
is added to the original link $U_{n,\mu}$ as
\begin{equation}
\Omega_{n,\mu}=(1-\alpha)U_{n,\mu}+\alpha'\,\Gamma_{n,\mu}\;.\label{eq:ape-sum}
\end{equation}
Here $\alpha'=\alpha/m$ and $m$ is the number of staples included
in the staple sum. Next $\Omega$, a general $N\times N$ matrix,
is projected back to $SU(N)$ as
\begin{equation}
V_{p}={\displaystyle \max_{V\in SU(3)}}\,\textrm{Re tr }(V\Omega^{\dagger})\,.
\label{eq:p-APE}
\end{equation}
In the following we will refer to this construction as projected-
or p-APE. Since no closed form for the derivative of the p-APE links
is known, they are difficult to use in molecular dynamics (MD) simulations.

Not long ago Peardon and Morningstar suggested a differentiable smearing
method~\cite{Morningstar:2003gk}. Their construction uses the staple
sum $\Gamma_{n,\mu}$ to define the differentiable $SU(N)$ stout
link as \begin{eqnarray}
V_{s} & = & e^{\rho S}U,\label{eq:def-stout}\\
S & = & \frac{1}{2}(\Gamma U^{\dagger}-U\Gamma^{\dagger})-\frac{1}{2N}\Tr(\Gamma U^{\dagger}-U\Gamma^{\dagger})\,.\label{eq:stout-def}\end{eqnarray}
 It is not obvious why the suggested form is a smearing at all beyond
the perturbative regime where $\Gamma U^{\dagger}\approx n\,\mathbb{I}$.
There the stout links are indeed identical to projected APE smeared
links with $\rho=\alpha/6$~\cite{Capitani:2006ni}. Nevertheless
stout smearing appears to work similarly to APE well beyond the perturbative
regime.

Here we consider a smeared link that is closer in spirit to the projected
APE links but it is differentiable and appropriate for MD simulations.
From the $N\times N$ general $\Omega$ matrix of Eq.~(\ref{eq:ape-sum})
we form a $U(N)$ unitary matrix as \begin{equation}
V_{n}=\Omega(\Omega^{\dagger}\Omega)^{-1/2}\,.\label{eq:W-def}\end{equation}
 Since $\Omega^{\dagger}\Omega$ is Hermitian and positive definite,
$(\Omega^{\dagger}\Omega)^{-1/2}$ is well defined, unless $\mathrm{det}\,\Omega=0$.
The smeared link $V_{n}$ is unitary but not in $SU(N)$, its determinant
in general is not one. The form in Eq.~(\ref{eq:W-def}) was first
used in Ref.~\cite{Liang:1992cz} to define smeared operators, while
in Ref.~\cite{Zanotti:2001yb} $V_{n}$ is divided by the cube-root
of its determinant to define an $SU(N)$ link.

One should note that there is no requirement that the smeared link
be an $SU(N)$ element, but in practice projecting the link back to
$SU(N)$ was found to be more effective in removing short scale fluctuations.
Here we will show that the $U(N)$ element $V_{n}$ link
is as effective as the projected smeared link. In the following we
will refer to the $V_{n}$ links defined in Eq.~(\ref{eq:W-def})
as normalized- or n-APE smearing. Since at the 1-loop perturbative
level neither the projection nor the normalization of the link plays
any role, the 1-loop perturbative properties of all three smearing
prescriptions are identical.

HYP smearing, as introduced in Ref.~\cite{Hasenfratz:2001hp}, consists
of three consecutive projected APE type smearing steps but the staple
sums at the higher level are constructed such that only links within
the hypercubes attached to the original link enter. The consecutive
smearing levels are constructed as

\begin{eqnarray}
V_{n,\mu} & = & \textrm{Proj}_{SU(3)}[(1-\alpha_{1})U_{n,\mu}+\frac{\alpha_{1}}{6}\sum_{\pm\nu\neq\mu}\widetilde{V}_{n,\nu;\mu}\widetilde{V}_{n+\hat{\nu},\mu;\nu}\widetilde{V}_{n+\hat{\mu},\nu;\mu}^{\dagger}]\,,\label{eq:HYP-def1}\\
\widetilde{V}_{n,\mu;\nu} & = & \textrm{Proj}_{SU(3)}[(1-\alpha_{2})U_{n,\mu}+\frac{\alpha_{2}}{4}\sum_{\pm\rho\neq\nu,\mu}\overline{V}_{n,\rho;\nu\,\mu}\overline{V}_{n+\hat{\rho},\mu;\rho\,\nu}\overline{V}_{n+\hat{\mu},\rho;\nu\,\mu}^{\dagger}]\,,\label{eq:HYP-def2}\\
\overline{V}_{n,\mu;\nu\,\rho} & = & \textrm{Proj}_{SU(3)}[(1-\alpha_{3})U_{n,\mu}+\frac{\alpha_{3}}{2}\sum_{\pm\eta\neq\rho,\nu,\mu}U_{n,\eta}U_{n+\hat{\eta},\mu}U_{n+\hat{\mu},\eta}^{\dagger}]\,.\end{eqnarray}
 The $U_{n,\mu}$ are the thin links from site $n$ in direction $\mu$,
the $V_{n,\mu}$ are the resulting HYP blocked fat links. The intermediate
fields $\widetilde{V}$ and $\overline{V}$ are constructed such that
the contributions to $V$ are restricted to the attached hyper-cube.
The indices after the semi-colon always indicate the directions excluded
from the sums. The three SU(3) projections make the HYP smeared configurations
very smooth while keeping the smearing within a hypercube ensures
that even short distance properties of the configurations are only
minimally distorted. While the main ingredient, the SU(3) projections,
make the HYP links difficult to use in dynamical simulations, any
of the above discussed differentiable smearings can be combined with
the HYP construction. In the following we will refer to the original
HYP links as p-HYP, to the normalized smearing as n-HYP and the stout
HYP construction as stout - or s-HYP. Again, at the 1-loop perturbative
level the three descriptions are identical~\cite{Capitani:2006ni}.

\subsection{Stout and n-APE smearing in SU(2)}

The two smearing prescriptions are easiest to compare for the gauge
group SU(2). The relevant quantity for both is $\Gamma U^{\dagger}$
which is a linear combination of SU(2) elements and can be written
as \begin{equation}
\Gamma U^{\dagger}=\omega_{0}\mathbb{I}+i\,\bar{\omega}\bar{\sigma,}\label{eq:su2}\end{equation}
 where $\omega_{0}$ and $\bar{\omega}=\hat{n}\omega$ are real. For
$SU(2)$ the traceless anti--Hermitian part of $\Gamma U^{\dagger}$,
$S$ in Eq.~(\ref{eq:stout-def}), is just $i\bar{\omega}\bar{\sigma}$
and we thus have \begin{equation}
V_{s}=e^{i\rho\bar{\omega}\bar{\sigma}}U=[\cos(\rho\omega)+i\,\sin(\rho\omega)\,\hat{n}\bar{\sigma}]U,\label{eq:stout-su2}\end{equation}
 while the APE link (\ref{eq:ape-sum}), normalized according to Eq.~(\ref{eq:W-def})
is \begin{eqnarray}
V_{n} & = & \Big[\frac{1+\xi\omega_{0}}{N_{0}}+i\frac{\xi\omega}{N_{0}}\hat{n}\bar{\sigma}\Big]U\,,\label{eq:nape-su2}\\
N_{0} & = & \sqrt{(1+\xi\omega_{0})^{2}+(\xi\omega)^{2}}\,,\label{eq:}\end{eqnarray}
 where $\xi=\alpha'/(1-\alpha)$. The stout link is independent of
$\omega_{0}$, the trace of $\Gamma U^{\dagger}$, and thus contains
less information about the original fields than the n-APE link.

Eqs.~(\ref{eq:stout-su2}) and (\ref{eq:nape-su2}) can nevertheless
be approximately identical if $\rho\omega,\;\xi\omega\ll1$, and $\omega_{0}$
can be replaced by its average value. According to Eq.~(\ref{eq:su2})
$\omega_{0}$ is related to the trace of the plaquettes around the
thin link $U$, so $\langle\omega_{0}\rangle=m\,\Tr(U_{{\rm {plaq}}})/N=m+O(\omega^{2})$
when $\omega\ll1$. These conditions are satisfied near the continuum
limit where fluctuations are suppressed and the gauge links are close
to the unit matrix. Then stout and n-APE links agree if $\xi\omega/N_{0}\approx\sin(\rho\omega)$,
or \begin{equation}
\rho=\frac{\xi}{1+\xi\langle\omega_{0}\rangle}=\frac{\alpha/m}{1-\alpha(1-\Tr(U_{{\rm {plaq}}})/N)}\,.\label{eq:stout-ape-corr}\end{equation}
 This relation agrees with the perturbatively expected form $\rho=\alpha/m$
if $\Tr(U_{{\rm plaq}})=N$. On typical MC configurations the plaquette
is considerably smaller than that, suggesting that even if stout and
n-APE smearing can be matched on MC configurations, the corresponding
stout parameter could be significantly different from the perturbatively
expected value. While the optimal parameter for APE smearing is largely
independent of the gauge coupling, this is not so for stout smearing.
On rough configurations where $\rho\omega$ is not small and $\omega_{0}$
cannot be replaced by its average, stout links could be very different
from n- or p-APE links and resemble little the form of Eq.~(\ref{eq:ape-sum}).

\subsection{Comparing projected, normalized and stout smearings}

Smearing reduces lattice artifacts by removing some of the non-physical
ultraviolet fluctuations of the gauge configurations. The effectiveness
of the smearing can be measured by the smoothness of the plaquette,
i.e. by the value of the average plaquette, and even more so by the
distribution of the smallest plaquette on finite volume configurations.
\begin{figure}
\includegraphics[scale=0.8]{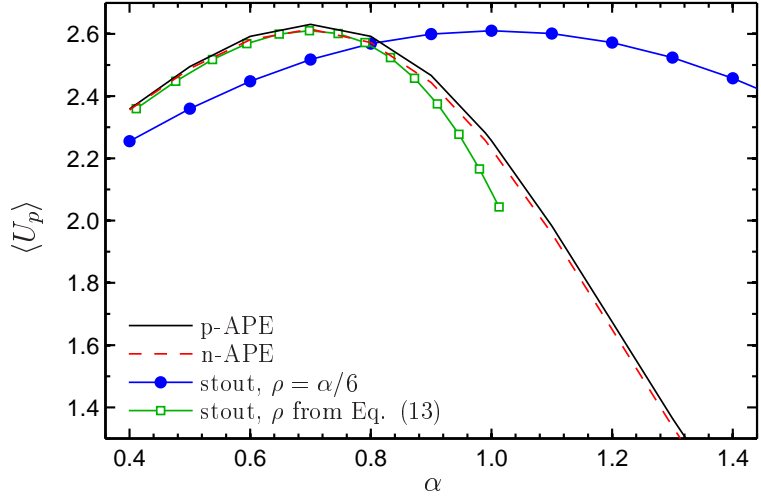}

\caption{The average plaquette on quenched $\beta=5.8$ configurations as
a function of the smearing parameter $\alpha$ after various single
level smearings. \label{fig:aver-plaq-ape} }
\end{figure}

The comparisons presented in this section are based on a set of 500
quenched $8^{4}$ lattices generated with the plaquette gauge action
at $\beta=5.8$, corresponding to a lattice spacing of 0.136~fm.
In Fig. \ref{fig:aver-plaq-ape} we show the average plaquette after
one level of p-APE, n-APE and stout smearing as a function of the
smearing parameter $\alpha$. The values measured after projected
and normalized APE smearing are nearly indistinguishable, predicting
the best smearing at about $\alpha=0.75$. Above this value the smearing
becomes unstable, the average plaquette drops even with only one level
of smearing. The stout smeared plaquette is plotted in two different
ways: once with the perturbatively predicted relation $\alpha=6\rho$,
and also with the relation based on the $SU(2)$ prediction of Eq.~(\ref{eq:stout-ape-corr}).
While the former parametrization leads to a very different result
than the APE smeared links, the latter one is surprisingly consistent
with those %
\footnote{It is usually assumed that the stout parameter must be tuned more
carefully than the APE parameter. This assumption is from Fig. 5 of
Ref.~\cite{Morningstar:2003gk} but one should note that in that
figure stout plaquettes are plotted against $\rho=\alpha/6$ while
n-APE smeared plaquettes are plotted as a function of $\alpha/6/(1-\alpha)$.
That way the plot covers the range $(0,\,3.0)$ for the stout links
but only $(0,\,0.75)$ for the n-APE links.%
}.

\begin{figure}
\includegraphics[scale=0.7]{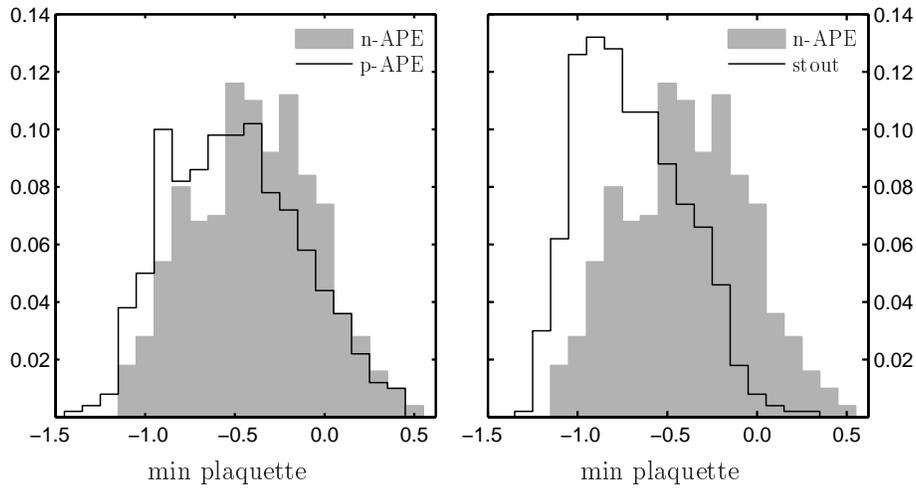}

\caption{The histograms show the distribution of the smallest smeared plaquette.
The left panel compares projected (lines) and normalized (shaded)
APE with $\alpha=0.75$ smearing parameter. The right panel compares
n-APE (shaded) with $\alpha=0.75$ and stout smearing with $6\rho=1.1$
(lines).\label{fig:APE-tail-distribution}}
\end{figure}

The most extreme fluctuations can be studied from the tail distribution
of the plaquette. Figure \ref{fig:APE-tail-distribution} shows the
histogram of the smallest plaquettes. The left panel compares p-APE
and n-APE smearings at the same $\alpha=0.75$ parameter value. It
is surprising how small the deviation is between the two smearings
even here when individual plaquettes are considered. If any difference
is observable, it is to the advantage of the n-APE smearing in the
sense that the latter produces slightly larger minimal plaquette values. The right
panel compares $\alpha=0.75$ n-APE and stout smearing at its optimal
value, $6\rho=1.1$. The difference is obvious, n-APE removes more
of the extreme fluctuations than stout smearing. Stout smearing with
$6\rho=0.75$ is considerably worse than n-APE smearing.

Next we consider HYP links based on the three different smearings.
In Ref.~\cite{Hasenfratz:2001hp} the projected-HYP parameters were
optimized by maximizing the smallest plaquette on a set of coarse
($\beta=5.7$) configurations. The optimal parameters found that way
($\alpha_{1}=0.75,$ $\alpha_{2}=0.6$ and $\alpha_{3}=0.3$) turned
out to be fairly independent of the gauge coupling and close to the
perturbative values that minimize taste violations for staggered fermions
($\alpha_{1}=0.875$ $\alpha_{2}=0.571$ and $\alpha_{3}=0.25$).
Since we found that n- and p-APE smearing are nearly identical numerically
and they are identical perturbatively, we expect that the same parameter
values are optimal for n-HYP as well. To optimize the stout-HYP parameters
we repeated the procedure of Ref.~\cite{Hasenfratz:2001hp}. We found
it difficult to identify an optimal parameter set, the sensitivity,
especially to the last parameter $\rho_{3}$, is weak compared to
statistical fluctuations. The best parameter values were large, even
larger than what one would predict based on Eq.~(\ref{eq:stout-ape-corr}),
and did not remove as many of the small plaquettes as n-HYP smearing.

In Table \ref{tab:Q-comparison-table}
we compare the average plaquette and the average of the minimum
plaquette values. 
In addition to n-HYP smearing with parameters $\alpha=(0.75,\,0.6,\,0.3)$
we consider 1, 2 and 3 levels of stout smearing with $6\rho=0.9$
smearing parameter, s-HYP smearing with parameters $6\rho=(0.85,\,0.75,\,0.35)$
and with parameters $6\rho=(1.2,\,1.0,\,0.4)$. The former s-HYP parameters
correspond to n-HYP parameters rescaled according to Eq.~(\ref{eq:stout-ape-corr}),
the latter one to the values found by optimizing the minimum plaquette
distribution. The average plaquette value does not always follow the
minimum plaquette. Based on the average plaquette one would expect
that 2 levels of stout smearing are about the same or better than
n-HYP. This expectation is false as we will show in Sect. \ref{sec:Numerical-tests}
. The minimum plaquette is a much better indicator of the quality
of smearing. That observable puts n-HYP close to 3 levels of stout
and considerable better than s-HYP even with the optimized b) parameter
set.

\begin{table}
\begin{tabular}{|c||c|c|}
\hline 
Smearing&
$\,\,\langle\Tr U_{p}\rangle\,\,$&
$\,\langle\Tr U_{p}^{{\rm min}}\rangle\,$\tabularnewline
\hline
\hline 
$1\times$ stout &
2.60&
-0.78(1)\tabularnewline
\hline 
$2\times$ stout &
2.84&
-0.16(2)\tabularnewline
\hline 
$3\times$ stout &
2.91&
0.46(3)\tabularnewline
\hline 
s-HYP$\,\,$a)&
2.80&
-0.39(2)\tabularnewline
\hline 
s-HYP$\,\,$b)&
2.64&
0.00(5) \tabularnewline
\hline 
n-HYP&
2.82&
0.38(3)\tabularnewline
\hline
\end{tabular}

\caption{Comparison of the plaquette and the minimum plaquette values. The
different smearings considered are: 1, 2 and 3 levels of stout smearing
with $6\rho=0.9$; stout-HYP a) with parameters $6\rho=(0.85,\,0.75,\,0.35)$
; b) with parameters $6\rho=(1.2,\,1.0,\,0.4)$; n-HYP with standard
HYP parameters $\alpha=(0.75,\,0.6,\,0.3)$; \label{tab:Q-comparison-table}}
\end{table}

To summarize our observations, we expect normalized-HYP to be as good
as projected-HYP with the same parameter values. The n-HYP parameters
do not have to be changed with the gauge coupling and perturbative
corrections are expected to be small. If stout-HYP smearing is used
in numerical simulations, the parameters will have to be tuned depending
on the gauge coupling. Stout-HYP smearing with parameters tuned that
way is effective in removing average fluctuations though it does not
work as well in removing the extreme fluctuations. At one-loop perturbation
theory all three smearings are identical, but since the optimal stout
parameters at large
or moderate lattice spacing are well above the
perturbative values, one expects larger perturbative corrections for
stout links. In dynamical updates the computational overhead for n-HYP
and stout-HYP is similar, therefore overall n-HYP appears to be a
better choice for simulations. In the following we describe the implementation
of n-HYP smearing in dynamical simulations.

\section{Force of the NHYP link\label{sec:Force-of-the}}

The equations of motion which are approximately solved in the molecular
dynamics evolution derive from ${\rm d}\mathcal{H}/{\rm d}\tau=0$,
where $\mathcal{H}=p^{2}/2+S_{f}+S_{g}$ is the molecular dynamics
Hamiltonian. The computation of the fermion contribution to the derivative
is subject of this section. We denote the part of the fermionic action
that depends on the smeared links by $S_{{\rm eff}}(V)$ and assume
its derivative $\Sigma_{\mu}$ with respect to the $V$ links has
already been performed. We now describe how to use the chain rule
to compute the derivative with respect to the thin links. In our discussion
we follow closely Ref.~\cite{Morningstar:2003gk}. We start out with
the derivative of $S_{{\rm eff}}$ with respect to the simulation
time parameter $\tau$ \begin{equation}
\frac{\textrm{d}}{\textrm{d}\tau}S_{{\rm eff}}=\mathrm{Re}\,\Tr\,\frac{\delta S_{{\rm eff}}}{\delta V_{\mu}}\frac{\textrm{dV}_{\mu}}{\textrm{d}\tau}\equiv\mathrm{Re}\,\Tr\,(\Sigma_{n,\mu}\dot{V}_{n,\mu})\;.\label{eq:force1}\end{equation}
 Here $\dot{V}=\textrm{d}V/\textrm{d}\tau$ refers to the derivative
with respect to the simulation time $\tau.$ Next we use the definition
of $V$ in terms of the thin links $U$ and the fat links $\widetilde{V}$
according to Eq.~(\ref{eq:HYP-def1}), with the projection replaced
by the normalization as given in Eq.~(\ref{eq:W-def}), to get \begin{eqnarray}
\textrm{Re\, tr\,(}\Sigma_{\mu}\dot{V_{\mu}}) & = & \textrm{Re tr}\Big[\Sigma_{\mu}^{(1)}\dot{U}_{\mu}+\widetilde{\Sigma}_{\nu;\mu}^{(1)}\dot{\widetilde{V}}_{\nu;\mu}\Big]\,,\label{eq:force2}\\
\Sigma_{n,\mu}^{(1)} & = & \Sigma_{n,\mu}\frac{\partial V_{n,\mu}}{\partial U_{n,\mu}}\,,\\
\widetilde{\Sigma}_{n,\nu;\mu}^{(1)} & = & \sum_{m,\rho}\Sigma_{m,\rho}\frac{\partial V_{m,\rho}}{\partial\widetilde{V}_{n,\nu;\mu}}\,,\end{eqnarray}
 where the sum over $m$ runs over all sites in the $\mu$--$\nu$
plaquettes attached to the link $(n,\mu)$ and $\rho$ can be either
$\mu$ or $\nu$. Next we express $\widetilde{V}_{\mu;\nu}$ in terms
of the thin links $U$ and smeared links $\overline{V}$ according
to Eq.~(\ref{eq:HYP-def2}), and continue this procedure until we
reach the level where only derivatives of the thin links are left
\begin{eqnarray}
\textrm{Re\, tr\,}{(\Sigma}_{\mu}\dot{V_{\mu}}) & = & \textrm{Re tr}\Big[(\Sigma_{\mu}^{(1)}+\Sigma_{\mu}^{(2)})\dot{U}_{\mu}+\overline{\Sigma}_{\rho;\nu,\mu}^{(2)}\dot{\overline{V}}_{\rho;\nu,\mu}\Big]\,\label{eq:force3}\\
 & = & \textrm{Re tr}\:\Big[(\Sigma_{\mu}^{(1)}+\Sigma_{\mu}^{(2)}+\Sigma_{\mu}^{(3)})\dot{U}_{\mu}\Big]\label{eq:force3a}\end{eqnarray}
 with \begin{eqnarray}
\Sigma_{\mu}^{(2)} & = & \widetilde{\Sigma}_{\nu;\mu}^{(1)}\frac{\partial\widetilde{V}_{\nu;\mu}}{\partial U_{\nu}}\,,\label{eq:force3b}\\
\overline{\Sigma}_{n,\rho;\nu,\mu}^{(2)} & = & \sum_{m,\alpha;\beta}\widetilde{\Sigma}_{\alpha;\beta}^{(1)}\frac{\partial\widetilde{V}_{m,\alpha;\beta}}{\partial\overline{V}_{n,\rho;\nu,\mu}}\,,\nonumber \\
\Sigma_{n,\mu}^{(3)} & = & \sum_{m,\alpha,\beta,\gamma}\overline{\Sigma}_{m,\alpha;\beta,\gamma}^{(2)}\frac{\partial\overline{V}_{m,\alpha;\beta,\gamma}}{\partial U_{n,\mu}}\,.\nonumber \end{eqnarray}
 Here we can finally identify $\Sigma_{\mu}^{(1)}+\Sigma_{\mu}^{(2)}+\Sigma_{\mu}^{(3)}=\delta S_{{\rm eff}}/\delta U_{\mu}$
as the fermionic force term.

Since the additional levels to Eq.~(\ref{eq:force2}) are very simple
modifications of the first level---only restricting the directions
the sum runs over---let us restrict the following discussion to the
first level. In terms of $\Omega$ defined in Eq.~(\ref{eq:ape-sum}),
the n-APE link is then given by $V_{\mu}=\Omega_{\mu}(\Omega_{\mu}^{\dagger}\Omega_{\mu})^{-1/2}$.
To compute the inverse square root of $Q=\Omega^{+}\Omega$, we employ
a method analogous to Morningstar and Peardon using the Cayley Hamilton
theorem. A non-singular $3\times3$ matrix $Q$ can always be written
as \begin{equation}
Q^{-1/2}=f_{0}\,\mathbb{I}+f_{1}Q+f_{2}Q^{2}\,,\label{eq:CH}\end{equation}
 where the scalars $f_{0}$, $f_{1}$, and $f_{2}$ are functions
of the traces of $Q$, $Q^{2}$ and $Q^{3}$ only. It is convenient
to define \begin{equation}
c_{0}={\textrm{tr\,}}Q\,\,\,\,;\,\,\,\, c_{1}={\textstyle \frac{1}{2}}\,{\textrm{tr\,}}Q^{2}\,\,\,\,;\,\,\,\, c_{2}={\textstyle \frac{1}{3}}\,{\textrm{tr\,}}Q^{3}\,.\end{equation}
 The details of the functional dependence of the $f_{i}$ on the $c_{j}$
is discussed in Sec.~\ref{sec:fb}.

To use the strategy indicated in Eq.~(\ref{eq:force2}), we apply
the chain rule until we are only left with derivatives of $U$ or
$V$, cycled to the right of the trace. For simplicity in the following
we drop the index $\mu$. \begin{eqnarray}
\textrm{Re\, tr}\,\Sigma\dot{V} & = & \textrm{Re\, tr}\Big(\Sigma\frac{\textrm{d}}{{\rm d\tau}}(\Omega Q^{-1/2})\Big)\nonumber \\
 & = & \textrm{Re\, tr}(Q^{-1/2}\Sigma\dot{\Omega})+{\textrm{tr}}(\Sigma\Omega)\dot{f}_{0}+{\textrm{tr}}(Q\Sigma\Omega)\dot{f}_{1}+{\textrm{tr}}(Q^{2}\Sigma\Omega)\dot{f}_{2}\label{eq:force4}\\
 &  & +f_{1}{\textrm{tr}}(\Sigma\Omega\dot{Q})+f_{2}{\textrm{tr}}((\Sigma\Omega Q+Q\Sigma\Omega)\dot{Q})\,.\nonumber \end{eqnarray}
Since the $f_i$ are scalar functions of the traces $c_n$ we get
\begin{equation}
\dot{f}_{i}=\sum_{n}\frac{\partial f_{i}}{\partial c_{n}}{\textrm{tr}}\big(Q^{n}\dot{Q}\big)\,.\end{equation}
 The computation of the derivatives $b_{ij}=\partial f_{i}/\partial c_{j}$
is described in the next section. Defining $B_{n}=b_{0n}+b_{1n}\, Q+b_{2n}\, Q^{2}$,
Eq.~(\ref{eq:force4}) leads to
\begin{equation}
\textrm{Re\, tr}(Q^{-1/2}\Sigma\dot{\Omega})+{\textrm{Re\, tr}}\left\{ \Big[\sum_{n}{\textrm{tr}}(B_{n}\Sigma\Omega)Q^{n}+f_{1}\Sigma\Omega+f_{2}(\Sigma\Omega Q+Q\Sigma\Omega)\Big]\dot{Q}\right\} \,.\label{eq:force5}\end{equation}
 Next, we define the sum in the square bracket as $A$ and use that
$Q=\Omega^{+}\Omega$ to get \begin{equation}
\textrm{Re tr}\left\{ (Q^{-1/2}\Sigma+A\Omega^{+}+A^{+}\Omega^{+})\dot{\Omega}\right\} \equiv{\textrm{Re}}\,\,{\textrm{tr}}(\Gamma\dot{\Omega})\end{equation}
 with $\Gamma=(A+A^{+})\Omega^{+}+Q^{-1/2}\Sigma$. To compute the
derivative of $\Omega$, we apply the chain rule again
\begin{eqnarray*}
\textrm{Re\, tr}(\Sigma_{n,\mu}\dot{V}_{n,\mu}) & = & \textrm{Re\, tr}(\Gamma_{n,\mu}\dot{\Omega}_{n,\mu})\\
 & = & \textrm{Re\, tr}\,\Gamma_{n,\mu}\Big[(1-\alpha)\dot{U}_{n,\mu}+\alpha'\sum_{\nu}\dot{\widetilde{V}}_{n,\nu;\mu}\widetilde{V}_{n+\hat{\nu},\mu;\nu}\widetilde{V}_{n+\hat{\nu}+\hat{\mu},\nu;\mu}^{+}\\[-2.5mm]
 &  &\qquad\qquad\ +\widetilde{V}_{n,\nu;\mu}\dot{\widetilde{V}}_{n+\hat{\nu},\mu;\nu}\widetilde{V}_{x+\hat{\nu}+\hat{\mu},\nu;\mu}^{+}+\widetilde{V}_{n,\nu;\mu}\widetilde{V}_{n+\hat{\nu},\mu;\nu}\dot{\widetilde{V}}_{n+\hat{\nu}+\hat{\mu},\nu;\mu}^{+}\Big]\,.
 \end{eqnarray*}
Now we can write down the final expression for $\Sigma^{(1)}$ .
First there is the {}``global'' contribution from the thin link
\begin{equation}
\Sigma_{n,\mu}^{(1)}=(1-\alpha)\Gamma_{n,\mu}\end{equation}
 and then there is the term that is multiplied with the derivatives
of the $\widetilde{V}$'s, which we have to collect from the various contributions
from neighboring sites\begin{eqnarray*}
\widetilde{\Sigma}_{n,\nu;\mu}^{(1)} & = & \alpha'\Big[\overline{V}_{n+\mu,\nu;\mu}\overline{V}_{n+\nu,\mu;\nu}^{\dagger}\Gamma_{n,\nu;\mu}+\overline{V}_{n+\mu,\nu;\mu}\Gamma_{n+\nu,\mu;\nu}\overline{V}_{n,\nu;\mu}^{\dagger}+\Gamma_{n+\mu,\nu;\mu}^{\dagger}\overline{V}_{n+\nu,\mu;\nu}^{\dagger}\overline{V}_{n,\nu;\mu}^{\dagger}\\
 &  & +(\nu\to-\nu)\Big]\,.\end{eqnarray*}
 The next term in the force expression, $\Sigma_{\mu}^{(2)}$, is
calculated the same way, by replacing $\Sigma_{\mu}$with $\widetilde{\Sigma}_{\mu;\nu}^{(1)}$
and $V_{\mu}$ with $\widetilde{V}_{\mu;\nu}$, and similarly for
$\Sigma_{\mu}^{(3)}$.

\subsection{Derivative of the $f$ constants\label{sec:fb}}

This section describes the computation of the Cayley-Hamilton constants
$f_{i}$ for the matrix $Q^{-1/2}$ and their derivatives with respect
to the traces of $Q^{n}$. The starting point is the definition in
Eq.~(\ref{eq:CH}). Since the matrix $Q=\Omega^{\dagger}\Omega$
is a positive, Hermitian matrix, it can be diagonalized with non-negative
eigenvalues $g_{i}$. Eq.~(\ref{eq:CH}) then translates into an
equation relating the eigenvalues to the coefficients $f$. \begin{equation}
\left(\begin{array}{ccc}
1 & g_{0} & g_{0}^{2}\\
1 & g_{1} & g_{1}^{2}\\
1 & g_{2} & g_{2}^{2}\end{array}\right)\left(\begin{array}{c}
f_{0}\\
f_{1}\\
f_{2}\end{array}\right)=\left(\begin{array}{c}
g_{0}^{-1/2}\\
g_{1}^{-1/2}\\
g_{2}^{-1/2}\end{array}\right)\label{eq:matrix-form}\end{equation}
 This equation has to be solved for $f$. Naturally, all expressions
are symmetric in the eigenvalues $g_{0}$, $g_{1}$ and $g_{2}$.
It turns out to be convenient to express the solution in terms of
the symmetric polynomials of the square roots of the eigenvalues $\sqrt{g_{i}}$\begin{equation}
u=\sqrt{g_{0}}+\sqrt{g_{1}}+\sqrt{g_{2}}\,\,\,\,;\,\,\
v=\sqrt{g_{0}g_{1}}+\sqrt{g_{0}g_{2}}+\sqrt{g_{1}g_{2}}\,\,\,\,;\,\,\
w=\sqrt{g_{0}g_{1}g_{2}}\,,\end{equation}
 such that we get for the coefficients $f$ the following results
\begin{eqnarray}
f_{0} & = & \frac{-w(u^{2}+v)+uv^{2}}{w(uv-w)}\nonumber \\
f_{1} & = & \frac{-w-u^{3}+2uv}{w(uv-w)}\\
f_{2} & = & \frac{u}{w(uv-w)}\,\,.\nonumber \end{eqnarray}
 To compute the symmetric polynomials, we need a closed formula of
the eigenvalues of $Q$ in terms of its traces (which are independent
of the basis) \[
c_{n}=\frac{1}{n+1}{\textrm{tr}}\, Q^{n+1}=\frac{1}{n+1}\sum_{i}g_{i}^{n+1}\,.\]
 This leads to a cubic equation whose solution is most easily expressed
in terms of \begin{eqnarray}
S=c_{1}/3-c_{0}^{2}/18\,\,\,\,;\,\,\,\, R=c_{2}/2-c_{0}c_{1}/3+c_{0}^{3}/27\,\,\,\,;\,\,\,\,\theta=\arccos\left(\frac{R}{S^{3/2}}\right)\end{eqnarray}
 with which the eigenvalues read for $n=0$, 1, 2\begin{eqnarray}
g_{n}=\frac{c_{0}}{3}+2\sqrt{S}\,\cos\left(\frac\theta3+(n-1)\frac{2\pi}{3}\right)\end{eqnarray}

Finally for their use in Eq.~(\ref{eq:force5}), we need to compute
the derivatives of the $f_{i}$ with respect to the traces $c_{j}$.
To this end, we use the chain rule and write \begin{equation}
B_{ij}=\frac{\partial f_{i}}{\partial c_{j}}=\sum_{k}\frac{\partial f_{i}}{\partial g_{k}}\frac{\partial g_{k}}{\partial c_{j}}\,.\end{equation}
 The matrix $\frac{\partial g_{k}}{\partial c_{j}}$ is the inverse
of the Vandermonde matrix $\frac{\partial c_{k}}{\partial g_{j}}=g_{j}^{k}$.
Factoring out the common denominator $d=2w^{3}(uv-w)^{3}$ we get
for the symmetric matrix $B=C/d$\begin{eqnarray*}
C_{00} & = & -w^{3}u^{6}+3vw^{3}u^{4}+3v^{4}wu^{4}-v^{6}u^{3}-4w^{4}u^{3}-12v^{3}w^{2}u^{3}\\
 &  & +16v^{2}w^{3}u^{2}+3v^{5}wu^{2}-8vw^{4}u-3v^{4}w^{2}u+w^{5}+v^{3}w^{3}\\
C_{01} & = & -w^{2}u^{7}-v^{2}wu^{6}+v^{4}u^{5}+6vw^{2}u^{5}-5w^{3}u^{4}-v^{3}wu^{4}-2v^{5}u^{3}\\
 &  & -6v^{2}w^{2}u^{3}+10vw^{3}u^{2}+6v^{4}wu^{2}-3w^{4}u-6v^{3}w^{2}u+2v^{2}w^{3}\\
C_{02} & = & w^{2}u^{5}+v^{2}wu^{4}-v^{4}u^{3}-4vw^{2}u^{3}+4w^{3}u^{2}+3v^{3}wu^{2}-3v^{2}w^{2}u+vw^{3}\\
C_{11} & = & -wu^{8}-v^{2}u^{7}+7vwu^{6}+4v^{3}u^{5}-5w^{2}u^{5}-16v^{2}wu^{4}-4v^{4}u^{3}+16vw^{2}u^{3}\\
 &  & -3w^{3}u^{2}+12v^{3}wu^{2}-12v^{2}w^{2}u+3vw^{3}\\
C_{12} & = & wu^{6}+v^{2}u^{5}-5vwu^{4}-2v^{3}u^{3}+4w^{2}u^{3}+6v^{2}wu^{2}-6vw^{2}u+w^{3}\\
C_{22} & = & -wu^{4}-v^{2}u^{3}+3vwu^{2}-3w^{2}u\,\,.\end{eqnarray*}
 Note that this expression is singular only for $w=0$, because $uv-w>0$
as long as one eigenvalue is non-zero. The pole in $w=\sqrt{g_{0}g_{1}g_{2}}$
corresponds to at least one zero eigenvalue of $Q$.

\section{Numerical tests\label{sec:Numerical-tests}}

The calculation of the fermionic force is considerably more involved
with HYP links than with stout links, but once the contribution from
the smearing is implemented, it can simply replace a stout smearing
force routine. Since in Ref. \cite{DeGrand:2004nq,DeGrand:2005af,Egri:2005cx,DeGrand:2006uy,DeGrand:2006ws,DeGrand:2006nv}
stout smearing was used in dynamical overlap simulations, we have
tested n-HYP smearing in the same set-up. We have also implemented
smearing in dynamical Wilson clover simulations. In the following
we briefly summarize our experience with n-HYP links, concentrating
mainly on algorithmic issues.

A common algorithmic concern, independent of the fermionic formulation,
is the potential occurrence of links with exactly zero determinant,
$\textrm{det}\Omega=0$. In such case the normalized smeared link
is ill-defined and the force term diverges. In our test runs we found
only once out of $10^{10}$ smeared link evaluations $\textrm{det}\Omega\approx10^{-8}$
and in single precision arithmetics that resulted in an exceptionally
large force term. The corresponding configuration was rejected and
the simulation continued without problem. In double precision even
this one occurrence could have been handled. The problem of $\textrm{det}\Omega\approx0$
might become much more severe at (even) coarser lattices but will disappear on
the way to the continuum.

\subsection{Overlap tests}

Smeared links are a common ingredient to chiral fermion simulations
because the cost of the Dirac operator application depends to a large
part on the spectral properties of the kernel operator it is constructed
from. To be specific, let us concentrate on Neuberger's overlap operator
\begin{equation}
D_{ov}=(R_{0}-\frac{m_{ov}}{2})\left[1+\gamma_{5}\epsilon(h(-R_{0}))\right]+m_{ov}\,,\end{equation}
 with $R_{0}$ the radius of the Ginsparg--Wilson circle, $m_{ov}$
the bare quark mass, $\epsilon$ the matrix sign function and $h=\gamma_{5}d$
the Hermitian Dirac kernel operator at negative mass shift $-R_{0}$.
$d$ is a Wilson like lattice Dirac operator, for our tests we take
the planar operator discussed in Refs.~\cite{DeGrand:2000tf,DeGrand:2004nq}.

Evaluating the action of the matrix sign function of $h$ on a vector
is the expensive part of overlap fermion simulations. The standard
technique is to compute the lowest few eigenmodes of $h$ explicitly
and use the spectral representation of the sign function for the corresponding
sub-space. For the rest of the spectrum, a polynomial or rational
approximation is used. In our test we use the Zolotarev rational approximation.
The approximated sign function therefore reads \begin{equation}
\epsilon(h)\approx h\sum_{i}\frac{b_{i}}{h^{2}+c_{i}}(1-\sum_{\lambda}P_{\lambda})+\sum_{\lambda}{\rm sign}\lambda\, P_{\lambda}\label{eq:zolo}\end{equation}
 with $P_{\lambda}$ the projector on the low-mode of $h(-R_{0})$
with eigenvalue $\lambda$. For each application of $D_{ov}$ on a
vector a multi-shift system with the kernel operator has to be solved.
Its condition number (and therefore the cost) decreases if the
region from which the modes are treated explicitly is increased. Firstly,
the lower bound of the Zolotarev approximation can be increased which
yields a larger minimal $c_{i}$. Secondly the smallest mode of $h^{2}$
which has not been projected is larger. Thus the condition number
of the whole system is smaller and it takes less iterations to solve
the system of linear equations. A lower density of modes at the origin
can therefore greatly reduce the cost of using the overlap operator.
This can be achieved by constructing the kernel operator $h$ from
smeared links.

To estimate how smearing in the kernel operator affects the cost of overlap
simulations, we compute one component of the overlap propagator at
mass $am_{ov}=0.03$ on 30 $12^{3}\times24$ dynamical clover configurations
described in Section~\ref{sub:clover}. On each configuration we
project out the lowest 10 eigenmodes of the kernel operator $h(-R_{0})$.
The number of iterations of the solver in the application of the sign
function is averaged over the whole computation of the propagator.
This gives the largest part of the cost of applying the overlap operator
in a realistic situation.

We compare kernel operators built from n-HYP links and stout
links with two and three levels of smearing. The stout smearing
parameter is set to $6\rho=0.9$ which is the value used in recent
calculations using dynamical overlap fermions~\cite{DeGrand:2006nv,DeGrand:2006uy},
while for n-HYP we use the standard HYP parameters. The results are
displayed in Table~\ref{tab:overlap}. The largest projected mode
is around 0.3 for both n-HYP and three levels of stout smearing whereas
it is roughly half that for 2 levels of stout smearing. Because the
smallest shift is much smaller than that, this also means that the
condition number of the former is a factor two smaller than for the
latter.

This is also reflected in the cost of applying the overlap operator.
Two levels of stout smearing is about twice as expensive
as either n-HYP or three iterated stout smearings. However, the
n-HYP smearing is more local than three levels of stout smearing and
also comes with smaller coefficients mixing the original links with
the staple.

\begin{table}
\begin{centering}\begin{tabular}{|c||c|c|c|}
\hline 
&
\ \ n-HYP\ \ &
$2\times$ stout &
$3\times$ stout \tabularnewline
\hline
\hline 
$\#_{h\times v}$ &
262(19) &
604(38) &
291(22) \tabularnewline
\hline 
$\langle|\lambda_{10}|\rangle$ &
0.31(1) &
0.16(1) &
0.28(1) \tabularnewline
\hline
\end{tabular}\par\end{centering}

\caption{\label{tab:overlap} The average number of applications of the kernel
operator per application of the overlap for different kinds of fat
links. We also give the average of the absolute value of the tenth
eigenvalue, the largest eigenvalue for which we use the spectral representation.}
\end{table}

\subsection{Wilson clover action tests\label{sub:clover}}

\begin{figure}
\includegraphics[scale=0.7]{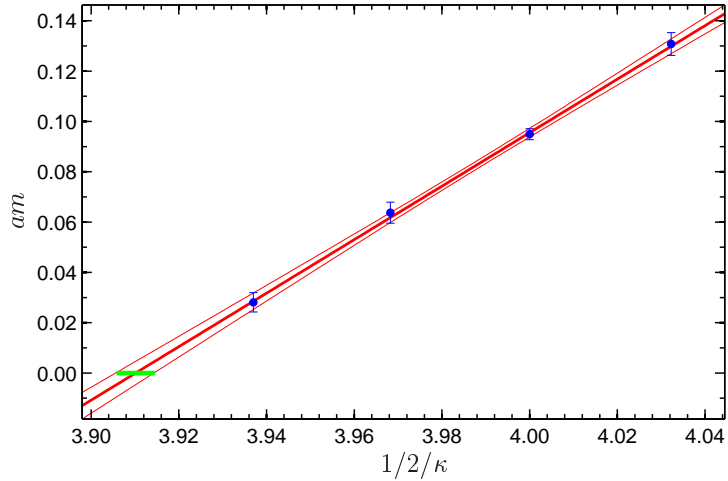}

\caption{The PCAC quark mass from simulations at $\beta=7.2$ on $8^{3}\!\times\!12$
lattices. The additive mass shift is only $am_{{\rm add}}=0.090(4)$.\label{fig:kappa}}
\end{figure}

We have implemented n-HYP smearing with two flavor ${\rm O}(a)$ improved Wilson fermions. For
the gauge action we use the L{\"u}scher-Weisz action and fix the tadpole
coefficient $u_{0}$ to be 0.875, the value that corresponds approximately
to our simulation values. Note that this choice affects the gauge
action only since the clover coefficient is left at its tree--level
value $c_{{\rm SW}}=1.0$; preliminary simulations indicated
that this is close to the value that minimizes the width of the spectral gap of the
Hermitian Dirac operator~\footnote{The tadpole improved value using the average
n-HYP plaquette of 2.81 would be $c_{{\rm SW}}=1.05$.}.
At $\beta=7.2$ the lattice spacing is around 0.13~fm and simulations
in smaller volumes (lattice size of $8^{3}\times12$) predict, from
the vanishing of the PCAC quark mass (see Fig.~\ref{fig:kappa}),
a critical hopping parameter of $\kappa_{c}=0.12787(14)$. This value
is surprisingly close to the one found in a quenched simulation with
p-HYP smearing at similar lattice spacing \cite{Capitani:2006ni}.
The additive mass shift is dramatically smaller for HYP links than
for thin link clover fermions, even with non--perturbative $c_{{\rm SW}}$.

The remaining results quoted in this section are obtained from
simulating a $12^{3}\times24$ lattice at $\beta=7.2$ and $\kappa=0.1266$.
We have accumulated 500 trajectories after thermalization and measured
eigenvalues of $D$ and $\gamma_{5}D$, n-HYP smeared Wilson loops,
as well as pseudoscalar correlators every 5 trajectories.

The three most expensive parts of the update are the calculation of
the fermionic force (including inversions), the gauge force and the
n-HYP blocking (including the n-HYP force term). Of the total CPU
time in these runs they consumed 75\%, 13\% and 11\%, respectively.
Thus, even with inexpensive fermion formulations such as Wilson the
computational overhead of the n-HYP blocking is negligible. One should
also note that the inversions of the Dirac operator are expected to
be significantly cheaper than in a comparable physical situation with
thin link clover fermions if the latter is possible at all.

A few remarks on the details of our simulation are in place:
Each trajectory was split in 25 steps using a Sexton-Weingarten
integrator and the same integrator on a finer time scale was also used
for the gauge force. This resulted in
an acceptance rate of 0.879(7).
On 32 nodes of a Myrinet cluster with 2GHz Xeon processors, 
one unit length trajectory took about 17 minutes to complete.

From fits to the static quark potential \cite{Hasenfratz:2001tw}
we extract the Sommer scale \cite{Sommer:1993ce} $r_{0}/a=3.903(25)$
and the string tension $a\sqrt{\sigma}=0.2897(26)$. The bare current
quark mass $am=0.0451(9)$ is in good agreement with the small volume
data shown in Fig. \ref{fig:kappa}, indicating small cutoff effects.
Assuming $r_{0}=0.5\,$fm we obtain a lattice spacing of $0.128(1)\,$fm
and bare current quark mass of $69.4(1.5)\,$MeV. We find a ratio
of pseudoscalar to vector meson mass of $0.57(3)$.

The behavior of the low-lying eigenmodes of the Dirac operator are
of particular interest if one wants to determine the degree of chiral
symmetry that is retained at finite lattice spacing. Also, the lowest eigenvalue
of the Hermitian Dirac operator $\gamma_{5}D$ is important for algorithmic
reasons as it determines the spectral gap and indicates the lowest
bare quark mass potentially accessible at a given lattice spacing
and volume~\cite{DelDebbio:2005qa}.

\begin{figure}
\includegraphics{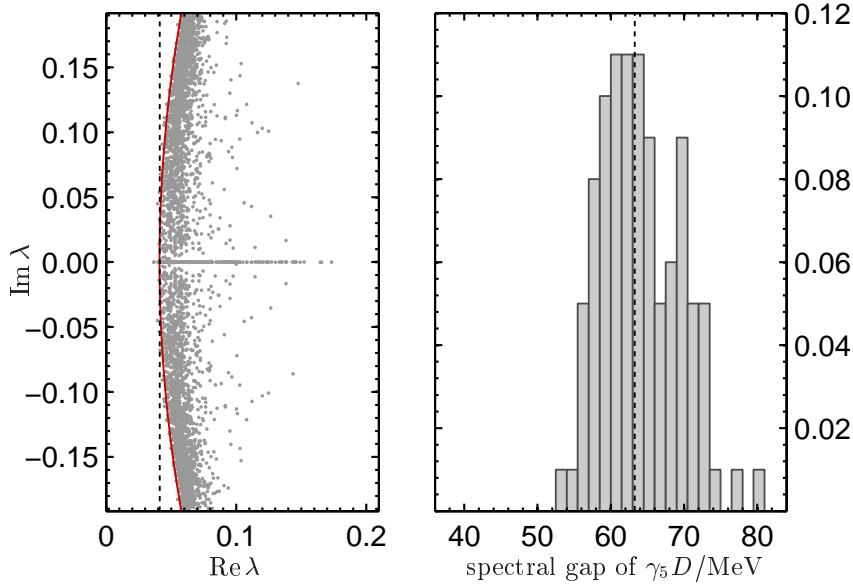}

\caption{The infrared spectrum of the n-HYP Wilson clover Dirac operator (left
panel), and the spectral gap as determined from the lowest eigenmode
of the Hermitian $\gamma_{5}D$ Dirac operator on the same configurations
(right panel). The dashed line in both plots indicates the median
of the spectral gap distribution. Data are from 100 $12^{3}\times24$
configurations with lattice spacing $a\approx0.13\,$fm and bare quark
mass $m\approx69\,$MeV. \label{fig:spec}}
\end{figure}

The left panel of Fig \ref{fig:spec} shows the infrared spectrum
(lowest 40 eigenmodes) of the n-HYP Dirac clover operator from 100
configurations. The complex spectrum has a rather well defined left
boundary that follows a circle with only a few real modes violating
that bound. This indicates that even at this coarse lattice spacing
much smaller quark masses can be reached without encountering exceptional
configurations. A similar plot is published in Ref.~\cite{Lang:2005jz}
showing the eigenmodes of the chirally improved CI Dirac operator
in two flavor dynamical simulations at similar lattice spacing and
volume, though about 50\% lighter quark masses. The spectrum in Fig.
\ref{fig:spec} compares well with that plot, showing similar widening
of the Ginsparg--Wilson circle for the two actions. 

A more direct measure of the accessible mass range is the spectral
gap, i.e. the distribution of the smallest magnitude eigenvalue of
the Hermitian Dirac operator $\gamma_{5}D$ \cite{DelDebbio:2005qa,DelDebbio:2007pz}.
This distribution is plotted on the right panel of Fig.~\ref{fig:spec}
with the median $\bar{\mu}=63.3(4)\,$ MeV marked by a dotted line.
The ratio of the median and the PCAC quark mass is indicative of the
renormalization factor $Z_{A}/(Z_{m}Z_{P})$ \cite{Capitani:2006ni,DelDebbio:2005qa}
and the value we obtain, 0.91, signals small perturbative corrections.

To facilitate comparison with similar distributions in Ref. \cite{DelDebbio:2005qa}
the data is plotted with the same bin size, $\Delta\mu=1.5\,$MeV.
The width of the distribution, defined as half the width of the shortest
interval that contains 68.3\% of the data, is $\sigma=5.5(6)\,$MeV.
Since the distribution in Fig. \ref{fig:spec} is quite asymmetric,
it is more physical to define the width as the interval to the left
of the median that contains 68.3\% of the data. This modified definition
gives $\sigma=4.6(6)\,$MeV. One expects that simulations at quark
masses of about $3\sigma$ are safe, which corresponds to $\simeq15\,$MeV at this
volume and lattice spacing. In Ref. \cite{DelDebbio:2005qa} it was
found that, at least for unimproved thin link Wilson fermions \cite{DelDebbio:2007pz},
the width of the spectral gap scales inversely with the square root
of the volume, $\sigma\sqrt{V}\approx1$. Assuming the same scaling
law in our case we find $\sigma\sqrt{V}\approx0.61-0.73$, depending
on the definition of the width. The decrease signals the improved
chiral properties of the smeared Dirac operator. The median $\bar{\mu}$
of the lowest mode of the Hermitian operator is also indicated in
the complex Dirac spectrum, where it is tangent to the circle that
bounds the spectrum.

\section{Conclusions}

We have successfully implemented and tested a gauge link smearing
scheme that inherits the good properties of HYP smearing (locality
and removal of dislocations) while still being suitable for MD based
algorithms. This is achieved by replacing the  projection
steps in the original HYP construction by normalizations to the corresponding
unitary group, thus allowing the calculation of the molecular dynamics
force for fermions coupled to the smeared links. We have tested the
n-HYP smearing with overlap fermions where we found that they can
be simulated as effectively as 3 level stout smeared fermions and
about twice as fast as 2-level stout smeared ones. We have also implemented
the smearing with Wilson -clover fermions. Our preliminary tests indicate
that light quarks, even as low as 15 MeV, can be simulated at $a\sim0.13\,$fm
lattices and volumes $aL\gtrsim1.6$ fm. In addition, the smoothness of the smeared
links speed up the inversion of the Dirac operator.

We have reported only preliminary results here. The volume, quark
mass, and lattice spacing dependence of Wilson clover simulations
with n-HYP links will be tested in the future.

\section{Acknowledgment}

At various stages of this project we have benefited from discussions with T. DeGrand, F. Niedermayer
and T. Kov\'acs.
We thank the computer center of DESY at Zeuthen for providing us with essential resources and support.
This research was partially supported by the US Dept. of Energy.
\bibliographystyle{apsrev}
\bibliography{lattice}

\end{document}